\documentclass[12pt,aps,nofootinbib]{revtex4}
\usepackage[english]{babel}
\usepackage{color}
 \usepackage{amsmath}
\usepackage{physics}
\usepackage[hidelinks]{hyperref}
\usepackage{diagbox}
\usepackage{multirow}
 
\usepackage{amsfonts}    
\usepackage{amssymb}
\usepackage{latexsym}
\usepackage{graphicx}
\usepackage{adjustbox}

\usepackage{float}

\newcommand{\la}{\lambda}

\newcommand{\rar}{\rightarrow}

\begin{document}

\title{Towards the Theory of the Yukawa Potential 
}

\author{J. C. del Valle}
\email{delvalle@correo.nucleares.unam.mx} 
\affiliation{Instituto de Ciencias Nucleares, Universidad Nacional Aut\'onoma de M\'exico, A. Postal 70-543 C. P. 04510, Ciudad de M\'exico, M\'exico.}
\author{D. J. Nader}
\email{daniel.nader@correo.nucleares.unam.mx} 
\affiliation{Instituto de Ciencias Nucleares, Universidad Nacional Aut\'onoma de M\'exico, A. Postal 70-543 C. P. 04510, Ciudad de M\'exico, M\'exico.}

\begin{abstract}

 Using three different approaches, Perturbation Theory (PT), the Lagrange Mesh Method (Lag-Mesh) 
 and  the Variational Method (VM), 
 we study  the low-lying states of  the Yukawa potential  $V(r)=-(\lambda/r)e^{-\alpha r}\,$. 
 First orders in PT in powers of $\alpha$ are calculated 
 in the framework of the Non-Linerization Procedure. 
 It is found that  the Pad\'e approximants to PT series 
 together with the Lag-Mesh
 provide highly accurate values  of the   energy
 and the positions of the radial nodes of the wave function. 
 The most accurate results, at present,
 of the critical screening parameters ($\alpha_c$) for some low-lying states and the first 
 coefficients in the expansion of the energy at $\alpha_c$ are presented.
 A locally-accurate and compact approximation for the eigenfunctions of the low-lying states
 for any $r\in [ 0,\infty)$ is discovered. This approximation used as a trial function in VM 
 eventually leads to  energies as 
 precise as those of PT and Lag-Mesh.
  Finally, a compact analytical expression for the energy as a function 
 of $\alpha$, that reproduce at least $6$ decimal digits in the entire physical
 range of $\alpha$, is found. 

\end{abstract}

\maketitle

\section{Introduction}
The Yukawa potential, sometimes called the screened Coulomb potential,
has a wide range of applications in many branches of physics.
Originally, Yukawa \cite{Yukawa} proposed this potential to describe the interaction between
a pair of  nucleons. However, it is often used  as a first approximation of the  interaction
between two dust   particles immersed in a plasma \cite{Plasma1,Plasma2,Plasma3}.
The  interaction
of dark matter particles through a Yukawa potential 
could explain the recently observed cores in dwarf galaxies \cite{Dark}. 
The Yukawa potential 
has also been employed to model the systems of colloidal particles in electrolytes \cite{Colloidal}. 

In the standard form the three-dimensional Yukawa potential is given by
\begin{equation}
\label{eq:Yukawa}
V(r)\ =\ -\lambda\,\frac{e^{-\alpha r}}{r}\ ,
\end{equation}
where $\alpha\geq0$ and $\lambda>0$ are  parameters and $r$ is the radial coordinate.
If $\alpha=0$, the Yukawa potential degenerates into
the Coulomb potential. When $\alpha \rightarrow\infty$, the Yukawa potential 
 vanishes: it describes a free motion.

In the non-relativistic  approximation the radial Schr\"odinger equation 
(in atomic units $\hbar=m=1$)  with the Yukawa potential (\ref{eq:Yukawa}) reads
\begin{align}
\label{eq:radial}
 \left[-\frac{1}{2}\,\frac{d^2}{dr^2}\  -\ \frac{1}{r}\,\frac{d}{dr}+\ \frac{l\,(l+1)}{2r^2}\ 
 -\ \lambda\,\frac{e^{-\alpha r}}{r}\right]\Psi_{n,l}\ =\ E_{n,l}(\lambda,\alpha)\,\Psi_{n,l}\ ,
\end{align}
  where   $n=0$, $1$, $...,$  is the principal quantum number and $l=0$, $1$, ..., $n-1,$ is
  the  angular momentum.
  It is well known that the Schr\"odinger equation (\ref{eq:radial}) with $\alpha\neq 0$
  is not exactly solvable and  only holds a finite number 
  of bound states.
  Naturally, this number depends on the parameters $\alpha$ and $\lambda$. 
  The scaling transformation $r\rightarrow \lambda^{-1} \,r$ in the  equation 
  (\ref{eq:radial}) allow us to remove the $\lambda$ dependence of the energy 
  since it is scaled as
  \begin{equation}
  E_{n,l}(\lambda,\alpha)\ =\ \lambda^2\,E_{n,l}(1,\alpha\,\lambda^{-1})\, .
  \end{equation}
  From now on we set $\lambda=1$.
  Consequently the number of bound states as well as the energy $E_{n,l}$ depends only on
  the value of $\alpha$.
  
  There is a certain critical value of $\alpha$,  denoted by $\alpha_{c}^{(n,l)}$,
  with the following  property: 
  for $\alpha<\alpha_c^{(n,l)}$ the state $(n,l)$ is bound being characterized by a normalized
  wave function, otherwise if $\alpha> \alpha_c^{(n,l)}$  the energy  of the  state $(n,l)$
 passes to the continuous regime and the corresponding state is no longer bound, i.e. 
 the wave function becomes non-normalizable.
 Hence, for the state $(n,l)\,$ the critical (screening) parameter 
 $\alpha_c^{(n,l)}$ is such that 
 \begin{equation}
 \label{Ecritical}
E_{n,l}\left(\alpha_c^{(n,l)}\right)\ =\ 0\ .
 \end{equation}
   There is a large number of papers
in the literature devoted to the calculation of the critical screening parameters
$\alpha_c^{(n,l)}$.
Let us list some of the approaches used for this purpose:  pseudo spectral method \cite{IJQC},
variational calculations \cite{lam,stubbins,variational},  perturbation theory \cite{Logarithmic},
finite scaling \cite{finite},
matrix propagation \cite{matrix}, direct solution of the Schr\"odinger equation \cite{Rogers,Weber}
and various numerical methods. 
At the moment there are certain discrepancies in the critical parameters $\alpha_{c}^{(n,l)}$, 
in general they differ in the sixth decimal digit. A brief historical discussion is 
given  in \cite{Weber} and complemented with \cite{katriel} about some estimates of the critical 
parameter of the ground state $\alpha_{c}^{(0,0)}$.
 
Following the results \cite{SIMON}, the energy $E_{n,l}$ can be 
represented as a series expansion at $\alpha_c^{(n,l)}$.
For  states with $l=0$ this expansion is
\begin{equation}
 \label{expansionE}
 E_{n,0}(\alpha)\ =\ \alpha^2\sum_{k=2}^{\infty}\beta_{k}^{(n,0)}\,\alpha^{-k}\left(\alpha_c^{(n,0)}-\alpha\right)^{k},
\end{equation}
while for states with $l\geq 1$ it reads 
\begin{equation}
 \label{expansionE2}
 E_{n,l}(\alpha)\ =\ \alpha^{2}\sum_{k=2}^{\infty}\beta_{k-1}^{(n,l)}\,\alpha^{-k/2}\left(\alpha_c^{(n,0)}-\alpha\right)^{k/2},
\end{equation}
where $\beta_{k}^{(n,0)}$ and $\beta_{k-1}^{(n,l\neq0)}$, $k=2,3,..$,  are real coefficients.
One of the aims of this study is to calculate the critical screening parameter $\alpha_c^{(n,l)}$
and 
 to find the first terms in the expansion (\ref{expansionE}) and (\ref{expansionE2}). 
We focus on the low-lying states with quantum numbers $n\leq5$ and $l\leq3$. 

Contrary to the situation of $\alpha_{c}^{(n,l)}$, the first coefficients
$\beta^{(n,l)}$ in (\ref{expansionE}) and (\ref{expansionE2}) are not known. 
The simplest way to estimate them   is to construct an interpolating function which  
 has the form of terminated expansions (\ref{expansionE}) and (\ref{expansionE2})
 and find coefficients by describing the energy in a close vicinity of $\alpha_c ^{(n,l)}$.
 In this manner the coefficients $\beta^{(n,l)}$  are free  
 parameters to be adjusted in a fit.  
 As we will discuss in Section III.C, $\beta_{2}^{(n,0)}$ and $\beta_{1}^{(n,l\neq0)}$ 
 are related to  the Hellmann-Feynman theorem, therefore 
 their calculation turns out to be appropriate to \textit{measure} the
 accuracy of an approximate wave function corresponding to $\Psi_{n,l}$ via an expectation value.
 
Our calculations of $\alpha_c^{(n,l)}$ 
 are carried out using two  approaches:  PT in powers of $\alpha$
and  Lag-Mesh.
The first one  is PT within the framework of the Non-Linearization Procedure
\cite{Turbiner1984}. 
Due to the probably divergent nature of this PT we will use a continued fraction representation
for the perturbation series \cite{Vrscay}.
The second one is Lag-Mesh \cite{Baye1} that, within its applicability domain,
is one  of the most accurate numerical
methods to solve the Schr\"odinger equation.

To compute $\beta_{k}^{(n,0)}$ and $\beta_{k-1}^{(n,l\neq0)}$, $k=2,3,...$, we proceed
with the interpolation procedure already described. 
 The calculation of the first terms in expansions (\ref{expansionE})
and (\ref{expansionE2}) plays
a fundamental role  in the construction of an analytical approximation to $E_{n,l}$ via
rational functions.
We present such approximations for some  states. Naturally, highly accurate energy
estimates are needed 
to construct accurate analytical approximations.
Therefore we complement the energy estimates that come from PT and Lag-Mesh 
with variational calculations. 
We construct a locally accurate and compact trial function  for the low-lying
states based on the interpolation 
of the series expansions of the wave function at $r=0$ and $r=\infty$.   
The Non-Linearization Procedure allows us to estimate the accuracy of the variational
calculations for the energy
and also for the trial function.

It is found that the three approaches (PT, Lag-Mesh and VM)  are  
also appropriate  to calculate the real radial 
nodes \{$r_0$\} of the wave function where it vanishes
 \begin{equation}
 \Psi_{n,l}(r_0)\ =\ 0\ .
\end{equation}
A precise calculation  of the real nodes (and more general, the nodal surface) 
is relevant in quantum mechanics. For instance, in  \cite{nodes} it is shown that 
the nodes plays an 
essential role for the variational  estimation of upper bounds of the energy of  excited states.
A similar situation occurs with fermionic systems studied with the diffusion
Monte Carlo method within 
the fixed-node approximation \cite{DMC}. 
 As we will present, the Non-Linearization Procedure is adequate to construct
 a PT for the factor of the wave function
 that defines  the radial  nodes. Again, since PT in powers of $\alpha$ is probably divergent,
 we use the continued fraction representation  to estimate $\{r_0\}$.
 It is found that the results of PT are in excellent agreement with those
 provided by the Lag-Mesh and VM.

\section{The methods}

\subsection{Perturbation Theory and its Summation }

\subsubsection{Non-Linerization Procedure}

This perturbative approach, sometimes called  Logarithmic PT, has shown to be a useful
tool in non-relativistic quantum
mechanics, see e.g.  \cite{Turbiner1984} and  \cite{Turbiner1984H}.  In particular,
it is a very efficient method 
to calculate  perturbation series. Let us give a brief review of this procedure by
 considering a generic spherically symmetric potential  $V(r)$.  

For convenience we use the following representation of the radial part of the wave function, 
\begin{equation}
\Psi_{n,l}(r)\ =\ r^lf_{n,l}(r)\,e^{-\Phi_{n,l}(r)}\ .
\label{eq:repexp}
\end{equation}
Note that if we restrict $e^{-\Phi_{n,l}(r)}\neq 0$ for  finite $r$ then the function $f_{n,l}(r)$ 
completely characterizes the nodal surface of any excited state $(n>1)$. Similar as in the one-dimensional  
case \cite{Turbiner1984},  we require  $f_{n,l}$ to be a polynomial of  degree ($n-l-1$)
with real (and non-negative) zeros.
Substituting (\ref{eq:repexp}) in the radial  Schr\"odinger equation (\ref{eq:radial}),
we obtain the non-linear
differential equation
\begin{equation}
\label{eq:riccati}
y_{n,l}'\ -\ y_{n,l}\left(y_{n,l}\ -\ \dfrac{2(l+1)}{r}\right)\ +\ \frac{  
2\,  \left(y_{n,l}\ -\ \dfrac{l+1}{r}\right)\,f'_{n,l}\ -\ f_{n,l}''}{ f_{n,l}}\ =\ 2\,(E_{n,l}-V)\ ,
\end{equation}
 where  $y_{n,l}(r)=\Phi'_{n,l}(r)$.
  The equation (\ref{eq:riccati}) is the starting point to develop PT. First, we assume
  that the   potential
  can be written as an expansion  in powers of some parameter $\gamma$, 
\begin{equation}
V(r;\gamma)\ =\ \sum_{k=0}^{\infty}V^{(k)}(r)\,\gamma^k\ ,
\label{eq:expansion}
\end{equation}
where $V^{(k)}$ are some given functions. 
  The  energy is also expanded in series of powers of $\gamma$, 
\begin{equation}
E_{n,l} \ =\ \sum_{k=0}^{\infty}  E^{(k)}_{n,l}\,\gamma^k\ , 
\label{eq:E}
\end{equation}
as well as the  functions $f_{n,l}$ and $y_{n,l}$,
\begin{equation}
 f_{n,l}(r) \ =\ \sum_{k=0}^{\infty}  f_{n,l}^{(k)}(r)\,\gamma^k\ , \quad\quad y_{n,l}(r) \ =\ \sum_{k=0}^{\infty}  y_{n,l}^{(k)}(r)\,\gamma^k\   . 
 \label{eq:fy}
\end{equation}

One can show that the linear differential equation  that determines
the $k$th corrections $E_{n,l}^{(k)}\,$, $f_{n,l}^{(k)}$ and $y_{n,l}^{(k)}$ is
\begin{eqnarray}
&(f^{(k)})''+\dfrac{2(l+1)}{r}(f^{(k)})'-2\sum_{i=0}^{k}(f^{(i)})'y^{(k-i)}\nonumber\\&+\sum_{i=0}^{k}\left\{f^{(i)}\left(\sum_{j=0}^{k-i}y^{(j)}y^{(k-i-j)}+2 \left( E^{(k-i)}-V^{(k-i)}\right)-\left(y^{(k-i)}\right)'-\dfrac{2(l+1)}{r}y^{(k-i)}\right)\right\}=& 0\nonumber\\
\label{eq:kcorrection}
\end{eqnarray}
where $\Phi^{(0)}_{n,l}(r)=\int y^{(0)}_{n,l}(r)\,dr\,$.
The boundary condition 
\begin{equation}
\label{boundary}	y_{n,l}^{(k)}\,e^{-\Phi^{(0)}_{n,l}}\Bigr|_{r\,=\,0,\,\{r_0\},\,\infty}\rar\ 0\ 
\end{equation}
must be  imposed \cite{Turbiner1984}.  
 For convenience we  have omitted
the labels $(n,l)$ in some expressions.  
Interestingly, the equation (\ref{eq:kcorrection}) can be further  developed  to obtain 
an integral representation 
for the $k$th energy  correction \cite{Turbiner1984}. One should note that
this approach does not require the previous knowledge 
of the entire spectrum of the unperturbed ($\gamma=0$) equation to construct perturbation series.

For the Yukawa potential (\ref{eq:Yukawa}) we set $\gamma=\alpha$ and consequently for the expansion (\ref{eq:expansion}) we have 
\begin{align}
\label{eq:unperturbed} 
V^{(k)}(r)\ &=\  \frac{(-r)^{k-1}}{k!}\ .
\end{align}
   With $\alpha=0$ the expansion (\ref{eq:expansion})
  corresponds to the  Coulomb potential. In this case the quantum numbers take the values $n=1,2,...$
  and $l=0,1,2,...,n-1$. The zeroth order corrections $E_{n,l}^{(0)}$, $f^{(0)}_{n,l}$  and $y^{(0)}_{n,l}$  are well known, 
\begin{equation}
E_{n,l}^{(0)}\ =\ -\frac{1}{2n^2}\ ,\quad\quad \quad f_{n,l}^{(0)}(r)\ =\ L_{n-l-1}^{2l+1}(r)\ ,\quad\quad \quad y_{n,l}^{(0)}\ =\ \frac{1}{n}\ , 
\end{equation}
where $L_{n-l-1}^{2l+1}(r)$ is the generalized Laguerre polynomial of degree $(n-l-1)$. For convenience we will denote 
\begin{equation}
L_{n-l-1}^{2l+1}(r)=\sum_{i=0}^{n-l-1}a_{n,l}^{(0)}[i]\,r^i
\end{equation} 
and choose $a_{n,l}^{(0)}[n-l-1]=1$ as  a normalization.

\subsubsection{Continued Fractions and Pad\'e Approximants}

For the ground state of the Yukawa potential, it is well known that PT (\ref{eq:E}) in powers of $\alpha$  is  divergent \cite{Logarithmic}. For excited states this PT is probably divergent.
This statement is also true for the series (\ref{eq:fy}).  Even  though perturbation series (\ref{eq:E}) are not
Stieldjes \cite{Vrscay}, the continued fraction representation can be used  to calculate accurately the energy
of a given state \cite{Lai,Vrscay}.
The basic idea is to assume that the energy $E_{n,l}$ has the representation
\begin{equation}
E_{n,l}(\alpha)\ =\ c_0\ +\ \cfrac{c_{1}\,\alpha}{1\ +\ \cfrac{c_{2}\,\alpha}{1\ +\ \cfrac{c_{3}\,\alpha}{1\ +\ \cdots}}}\ ,
\end{equation}
where $c_J$, with $J=0,1,...$,  are  coefficients to be determined. For convenience we omit the label ($n,l$) in the coefficients $c_J$. 
 In  practice, the continued fraction is truncated by setting $c_{J}=0$ if $J>M$, where $M$ is some  positive integer. As a result of this truncation, a Pad\'e approximant of the form  $P^{ \left \lceil{M/2}\right \rceil }_{\left \lfloor{M/2}\right \rfloor}(\alpha)$ emerges,
 \begin{equation}
 \label{eq:pades}
P^{ \left \lceil{M/2}\right \rceil }_{\left \lfloor{M/2}\right \rfloor}(\alpha)\ 
=\ \frac{\sum_{k=0}^{\left \lceil{M/2}\right \rceil}C_k\,\alpha^k}{\sum_{k=0}^{\left \lfloor{M/2}
\right \rfloor}D_k\,\alpha^k}\,,
 \end{equation}
where$ \left\lceil{\ }\right \rceil$ and $\left \lfloor{\ }\right \rfloor$ denote the ceiling and floor functions,
respectively. The coefficients $C_k$ and $D_k$ are determined by demanding 
 \begin{equation}
\sum_{k=0}^{M}E_{n,l}^{(k)}\,\alpha^{k}\ -\ P^{ \left \lceil{M/2}\right \rceil }_{\left \lfloor{M/2}\right \rfloor}(\alpha)\ =\ O\left(\alpha^{M+1}\right)\ .
 \end{equation}
 Once the Pad\'e approximant is completely  determined it is used to calculate the energy $E_{n,l}(\alpha)$,   the  value of the critical parameter $\alpha_c^{(n.l)}$ and the coefficients  and $\beta^{(n,0)}_{k}\,$ and $\beta^{(n,l\neq0)}_{k-1}\,$.
 Additionally,
 we will show that if the continued fraction representation  is 
 assumed for the perturbation series of $f_{n,l}\,$, see (\ref{eq:fy}),  
 we can calculate the position of the nodes with high accuracy via Pad\'e approximants.

\subsection{The Lagrange Mesh Method }
The Lag-Mesh has shown to be a simple and very accurate method to solve the 
Schr\"odinger equation, see  \cite{Baye1,Baye2}.
Essentially, the wave function is expanded
in terms of the Lagrange functions while  the  Gauss quadrature is used
to calculate approximately the matrix elements of the Hamiltonian.
Once the matrix elements are known, we proceed to calculate 
 the eigenvalues and the corresponding eigenfunctions.  In the next two Subsections
 we give a brief review of the method
nevertheless it is described in full detail in \cite{Baye1,Baye2,Baye3}.

\subsubsection{The Lagrange Functions and the Gauss Quadrature}
 
 A   mesh of dimension $N$ involves  $N$ real zeros $r_{i=1,...,N}$ of 
 a particular orthogonal polynomial $P_N(r)$ of degree $N$. 
Given 
the values of a function $F(r)$ at $r_i\,$,
the polynomial  of minimal grade $(N-1)$, denoted by $L_{N-1}(r)$,  which 
interpolates the function $F(r)$ is  of the form
\begin{equation}
\label{Lagrange} 
 L_{N-1}(r)\ = \ \sum_{i=1}^NF(r_i)\,f_i(r)\ ,
\end{equation}
where the Lagrange functions $f_i(r)$ are defined by
\begin{equation}
\label{polinomio}
 f_i(r)\ =\ \frac{P_N(r)}{(r-r_i)\,P_N^{\prime}(r_i)}\ .
\end{equation}
Since $r_i$ are the roots of $P_N(r)$ the Lagrange functions satisfy the property $f_i(r_j)=\delta_{ij}$,
and therefore   $L_{N-1}(r_i)=F(r_i)$.
The integral of the function $F(r)$ in the domain $[a,b]$ can be  approximated using the 
Gauss Quadrature as follows
\begin{equation}
\label{eq:quadrature}
\int_{a}^{b}F(r)\,dr\ \approx\ \sum_i^N \la_i\,F(r_i)\,,
\end{equation}
where $\la_i=\int_{a}^{b}f_i(r)dr$ are  the associated  weights. The Gauss quadrature
provides  high accuracy on the integrals except when the function $F(r)$ contains
singularities or discontinuities \cite{Baye1}.

\subsubsection{The Lag-Mesh in Quantum Mechanics}
For spherically symmetric potentials it is convenient to transform the radial 
Schr\"odinger equation (\ref{eq:radial}) into its one-dimensional counterpart. 
If we assume a wave function of the form $\Psi_{n,l}(r)=r^{-1}u_{n,l}(r)$, then the function $u_{n,l}(r)$
satisfies
\begin{equation}
\label{radial1}
 \left[-\frac{1}{2}\,\frac{d^2}{dr^2}\ +\ U(r)\right]u_{n,l}\ =\ E_{n,l}\,u_{n,l}\ ,
\end{equation}
with the effective potential $U(r)$ given by 
\begin{equation}
U(r)\ =\ V(r)\ +\  \frac{l\,(l+1)}{2\,r^2}\ .
\end{equation}
Now, we consider a wave function  $u_{n,l}$  given as an  expansion 
\begin{equation}
\label{eq:ulag}
 u_{n,l}(r)\ =\ \sum_{i=1}^N c_i\,\hat{f}_i(r)\ , 
 \end{equation}
 where $c_i$ are  coefficients and $\hat{f}_i(r)$  are
  the  regularized Lagrange functions
 \begin{equation}
 \label{regularized}
 \hat{f}_i(r)\ =\ \lambda_i^{-1/2}\,\frac{r}{r_i}w(r)^{1/2}\,f(r)\ ,
 \end{equation}
 here $w(r)$ is the weight function associated to the orthogonal polynomial $P_N(r)$.
One can immediately notice that
in this  representation the function $u_{n,l}(r)$ always vanishes at the origin. 
The other boundary condition, $u(r)\rar0$ as $r\rar\infty$, is satisfied by choosing the appropriate  
polynomial $P_N(r)$. Since we are interested in the domain $r\in[0,\infty)$ the 
Laguerre mesh is adequate. Therefore, we set $P_N(r)=L_N^{0}(r)$, where $L_N^{0}(r)$ is the $N$th Laguerre polynomial 
and the weight function is $w(r)=e^{-r}$.  

 The coefficients $c_i$ are determined by the secular equation related  to (\ref{radial1}),
\begin{equation}
\label{secular}
\sum_{j=1}^{N} \left\{\,T_{ij}\ +\ U_{ij}\ \right\}c_j\ =E\,c_i \ ,
\end{equation}
where the matrix elements are 
\begin{equation} 
T_{ij}\ =\ -\,\frac{1}{2}\int \hat{f}_i(r)\,\frac{d^2}{dr^2}\,\hat{f}_j(r)\,dr\ ,\quad\quad\quad U_{ij}\ =
\ \int \hat{f}_i(r)\,U(r)\,\hat{f}_j(r)\,dr\ .
\end{equation}

With the Gauss quadrature the matrix elements of the effective potential $U_{ij}$ are found 
\begin{eqnarray}
\label{matrixV}
U_{ij}&=& 
U(r_i)\,\delta_{ij}\ =\ \left(V(r_i)\ +\ \frac{l\,(l\ +\ 1)}{2\,r_i^2}\right)\delta_{ij}\ .
\end{eqnarray} 
Therefore, we can see that the  matrix representation of $U$ is diagonal in the Gauss approximation.
For the  kinetic matrix elements $T_{ij}\,$,   the discrete variable 
representation \cite{Szalay}
of the operator $d^2/dr^2$ is useful to obtain the elements in closed form  within the Gauss quadrature,
\begin{align}
T_{ii}\ =&\ \frac{4\ +\ (4N\ +\ 2)\,r_i\ -\ r_i^2}{24\,r_i^2}\ ,\quad\quad i= j\ ,\\
T_{ij}\ =&\ \frac{(-1)^{i-j}\,(r_i\ +\ r_j)}{2\,(r_i\, r_j)^{1/2}\,(r_i \ -\ r_j)^2}\ ,\quad\quad i\neq j\ .
\end{align}
It is worth mentioning that the use of the regularized   Lagrange functions circumvents the error of the Gauss quadrature induced
by the  singularity of the Yukawa potential at $r=0$.
Once  all the matrix elements are known, we proceed to solve the secular equation (\ref{secular}) for the Yukawa potential (\ref{eq:Yukawa}). 
From the solution of the secular equation we obtain the first $N$ approximate wave functions and their corresponding energies
for a fixed angular momentum $l$ and parameter $\alpha$.
Interestingly, the expectation value of any function ${g}(r)$  
can be computed (in the Gauss quadrature approximation) as follows 
\begin{eqnarray}
\label{matrixV}
\langle \Psi_{n,l} |g(r)|\Psi_{n,l}\rangle
\ =\ \sum_{i=1}^{N} |c_i|^2\,g(r_i)\  .
\end{eqnarray}

\subsection{Trial Functions}
 
In order to design a trial function for the state $(n,l)$
we will follow the approach presented in \cite{TURBINER2005}. In the latter, a
trial function was constructed for the one-dimensional quartic anharmonic and double-well  potentials which leaded to 
the most accurate variational  energy estimates of the ground state.
The basic idea is to construct a minimal interpolation between the expansions of 
the wave function at $r=0$ and $r=\infty$.

We begin the construction by considering the ground state wave function $\Psi_{1,0}$ in the 
representation  (\ref{eq:repexp}),
\begin{equation}
\label{eq:exprep}
\Psi_{1,0}(r)\ =\ e^{-\Phi_{1,0}(r)}\ .
\end{equation}
Using (\ref{eq:radial}), it is straightforward to show that the function $y_{1,0}=\Phi_{1,0}'$ satisfies a non-linear differential equation, namely
\begin{equation}
\Phi_{1,0}''-\Phi'_{1,0}\left(\Phi'_{1,0}-\frac{2}{r}\right)=2\left(E_{1,0}+\frac{e^{-\alpha r}}{r}\right)\ .
\end{equation}
From this equation one can construct the series expansions 
of the function   $\Phi_{1,0}$  at $r=0$ and $r=\infty$. These expansions read
\begin{equation}
\label{eq:phir0}
\Phi_{1,0}(r)\ =\ \ r\ +\ \frac{1}{6}(1-2\alpha+2E_{1,0})\,r^2\ +\ \frac{1}{36}(2-4\alpha+3\alpha^2+4E_{1,0})\,r^3\ +\ \ldots\ ,
\end{equation}
\begin{equation}
\label{eq:phirinfty}
\Phi_{1,0}(r)\ =\ \sqrt{-2E_{1,0}}\,r\ +\ \ln r\ -\ \frac{4}{\alpha(\alpha+2\sqrt{-2E_{1,0}})}\,\frac{e^{-\alpha r}}{ r}\ -\ \frac{4}{\alpha(\alpha+2\sqrt{-2E_{1,0}})^2}\frac{e^{-\alpha r}}{r^2}\ +\ \ldots\ ,
\end{equation}
for $r=0$ and $r=\infty$, respectively.
 In the limit $\alpha\rar0$ 
 both expansions are truncated and they coincide 
\begin{equation}
\Phi_{1,0}(r)\ =\ r\ .
\end{equation}
Together with (\ref{eq:exprep}), this  is nothing but the ground state wave function of the hydrogen atom.
 One of the simplest  interpolations between (\ref{eq:phir0}) and (\ref{eq:phirinfty}) is given by 
 \begin{equation}
 \label{eq:phase}
\Phi_{1,0}^{(t)}\ =\ r\left(\frac{a_{1,0}\,r\ +\ b_{1,0}\, e^{-\alpha r}\ +\ c_{1,0}\,e^{-2 \alpha r } }{d_{1,0}\,r\ +\ e^{-a r}}\right)\ +\ \log (d_{1,0}\, r\ +\ e^{-\alpha r})
 \end{equation}
 where $\{a_{1,0},b_{1,0},c_{1,0},d_{1,0}\}$ are variational parameters. In this manner, our trial function for the ground state is 
  \begin{equation}
 \label{tfgs}
 \Psi_{1,0}^{(t)}(r)\ =\ e^{-\Phi_{1,0}^{(t)}(r)}\  .
 \end{equation}
For excited states we  use $\Psi_{1,0}^{(t)}$ 
as a building block to construct the trial function of the state $(n,l)$
 \begin{equation}
 \label{tfes}
 \Psi_{n,l}^{(t)}(r)\ =\ f_{n,l}^{(t)}(r)\,e^{-\Phi_{n,l}^{(t)}(r)}\ ,\quad\quad f_{n,l}^{(t)}(r)\ =r^l\,P^{(t)}_{n,l}(r)\ ,
 \end{equation}
 where $P^{(t)}_{n,l}(r)$ is a polynomial of degree $n-l-1$.  Here $\Phi_{n,l}^{(t)}$ has the same functional
 structure as (\ref{eq:phase}) with a different set of parameters $\{a_{n,l},b_{n,l},c_{n,l},d_{n,l}\}$. 
 We impose the constraint that the trial function $\Psi_{n,l}^{(t)}$ must be 
 orthogonal to the functions $\Psi_{n-1,l}^{(t)}$, $\Psi_{n-2,l}^{(t)}$, $...$, $\Psi_{l+1,l}^{(t)}$. 
 This constraint fixes the value of some parameters of $\Psi_{n,l}^{(t)}$. The remaining free parameters are
 taken as variational parameters. This parameters are adjusted 
 such that the expectation value of the radial Hamiltonian 
 \begin{equation}
 \label{eq:radialoperator}
 \hat{h}\ =\ -\frac{1}{2}\,\frac{d^2}{dr^2}\  -\ \frac{1}{r}\,\frac{d}{dr}\ +\ \frac{l\,(l+1)}{2r^2}\ \ -\frac{e^{-\alpha r}}{r}\,,
 \end{equation}
 is minimal.
 
 Interestingly, there  is a connection between PT and VM \cite{Turbiner1984} that we explain briefly.
 Any  potential $V(r)$ can always be rewritten as 
 \begin{equation}
 \label{eq:rewritten}
 V(r)\ =\ V_0(r)\ +\ \gamma\,\left(V(r)\ - \ V_0(r)\right)\ ,\quad\quad \gamma=1\ ,
 \end{equation}
 where $\gamma$ is formal parameter and $V_0$ is the potential for which the trial function $\Psi^{(t)}_{n,l}$ is
 the exact solution of the Schr\"odinger  equation
\begin{align}
 \left[-\frac{1}{2}\,\frac{d^2}{dr^2}\  -\ \frac{1}{r}\,\frac{d}{dr}\ +\ \frac{l\,(l+1)}{2r^2}\  +\  V_0(r)\right]\Psi^{(t)}_{n,l}\ =\ E^{(0)}_{n,l}\,\Psi^{(t)}_{n,l}\ .
\end{align}
 In equation (\ref{eq:rewritten}) 
 \begin{equation}
 V^{(1)}(r)\ =\ V(r)\ -\ V_0(r)
 \end{equation}
plays the role  of the perturbation potential.
Consequently, the variational  energy 
corresponds to the first two terms of a perturbative series,  
 \begin{equation}
 \frac{\int_0^{\infty}\Psi^{(t)}_{n,l}\,\hat{h}\, \Psi^{(t)}_{n,l}\,r^2\,dr}{\int_0^{\infty}(\Psi^{(t)}_{n,l})^2\,r^2\,dr}\ =\ E^{(0)}_{n,l}\ +\ \gamma\,E^{(1)}_{n,l}\ ,\quad\quad\quad E_{n,l}^{(1)}\ =\ \frac{\int_0^{\infty}\Psi^{(t)}_{n,l}\,V^{(1)}\, \Psi^{(t)}_{n,l}\,r^2\,dr}{\int_0^{\infty}(\Psi^{(t)}_{n,l})^2\,r^2\,dr}\ .
 \end{equation}
 Since the Non-Linearization Procedure only requires as entry the unperturbed wave function, 
 we can take the trial function $\Psi^{(t)}_{n,l}$ as an unperturbed wave function and then develop PT in order to
 construct higher order corrections for  $E_{n,l}^{(0)}$,
 $f_{n,l}^{(t)}$ and $\Phi_{n,l}^{(t)}$. In this manner we can estimate the accuracy of our calculations by means of the perturbative corrections.
 As a consequence we can define (and calculate), for example 
\begin{equation}
E_{n,l;1} \ = \ E^{(0)}_{n,l}\ +\ E^{(1)}_{n,l} 
\end{equation} 
 and
 \begin{equation}
 E_{n,l;2}\ =\ E^{(0)}_{n,l}\ +\ E^{(1)}_{n,l}\ +\ E^{(2)}_{n,l}\,,
 \end{equation}
 which correspond to the first and second order approximations to the energy. In general we can define the sum of the corrections
 \begin{equation}
 \label{notation}
 E_{n,l;i}\ =\ \sum_{k=0}^{i}E_{n,l}^{(k)}
 \end{equation}
 as the $i$-th approximation to the exact energy $E_{n,l}$. 
In particular,  if the trial function is chosen appropriately a convergent PT occurs \cite{Turbiner1984},
 \begin{equation}
 \label{eq:limit}
\lim_{i\rar\infty}E_{n,l;i}\ =\ E_{n,l}\ .
 \end{equation}
Similar approximations can be defined for the functions  $f_{n,l}^{(t)}$ and $\Phi_{n,l}^{(t)}$. However, if the exact position of the node is known and this information is codified in the trial function, no correction of $f_{n,l}^{(t)}$ should be calculated. This situation occurs for the states with quantum numbers ($l+1,l$) which only have a node of order $l$ at $r=0$.

\section{Results}

\subsection{Some Remarks on PT}
We begin by presenting some results  about the realization of PT in the framework of the Non-Linearization Procedure. 

The first order corrections always vanish, $f^{(1)}_{n,l}=y_{n,l}^{(1)}=0$. We found that the corrections $f_{n,l}^{(k)}(r)$ and $y_{n,l}^{(k)}(r)$  are both polynomials in $r$ of the form 
\begin{equation}
\label{eq:cf}
f_{n,l}^{(k)}(r)\ =\ \sum_{i=0}^{n-l-2}a_{n,l}^{(k)}[i]\,r^i\ ,\quad\quad k>1\ ,
\end{equation}
and
\begin{equation}
\label{eq:cy}
y^{(k)}_{n,l}(r)\ =\ \sum_{i=0}^{k-1}b_{n,l}^{(k)}[i]\,r^i\ ,\quad\quad k>1\ ,
\end{equation}
respectively.  The coefficients $a_{n,l}^{(k)}[i]$ and $b_{n,l}^{(k)}[i]$ are always real and rational numbers. 
Since the corrections are polynomials the realization of PT is an algebraic and  iterative procedure.
More precisely, using expressions (\ref{eq:cf}) and (\ref{eq:cy}), the differential equation (\ref{eq:kcorrection}) transforms into 
an algebraic equation for  $a_{n,l}^{(k)}[i]\,$, $b_{n,l}^{(k)}[i]$ and the energy correction $E_{n,l}^{(k)}\,$. The expansion of $E_{n,l}$ in powers of $\alpha$ is an alternating series and the coefficients $E_{n,l}^{(k)}$, $k=0,1,...$, are also rational numbers.
 The algebraic nature of the realization of the PT allowed us to calculate high orders in PT by using the 
 software \textit{Mathematica}. The Pad\'e approximants to the series can be easily constructed using also this software.
 For some selected states  the first 10 corrections $E^{(k)}_{n,l}$  are presented in the Table \ref{Pcoeff}. However, for the states considered, we were able to calculate exactly the first 400 perturbative corrections in (\ref{eq:E}) and (\ref{eq:fy}). For example, for the energy of the ground state, some high-order coefficients (rounded) are: $E_{1,0}^{(100)}=-7.6658\times10^{84}$, $E_{1,0}^{(200)}=-1.1169\times 10^{211}$, $E_{1,0}^{(300)}=-5.1980\times10^{354}$ and $E_{1,0}^{(400)}-1.1292\times10^{510}$.
 
From the polynomial structure of the corrections $f_{n,l}^{(k)}$ we corroborate that the function $f_{n,l}(r)$ is  also a polynomial, 
\begin{equation}
f_{n,l}(r)\ = r^l\,\sum_{k=0}^{n-l-1}A_{n,l}[k]\,r^{k}\ 
\label{eq:nodal},
\end{equation}
where
\begin{equation}
 A_{n,l}[n-l-1]\ =\ 1\ 
 \end{equation}
 and
 \begin{equation}
  A_{n,l}[k]\ =\ \sum_{i=0}^{\infty}a_{n,l}^{(k)}[i]\,\alpha^i\quad \quad\text{for $k<n-l-1$}\ .
\label{eq:coeff}
\end{equation}
Since both series,  for the energy $(\ref{eq:E})$ and for coefficients (\ref{eq:coeff}) of $f_{n,l}(r)$,  are probably divergent, 
we used a continued fraction representation  (and hence Pad\'e approximants) for the  summation.
Therefore, once the coefficients (\ref{eq:coeff}) are known, the radial nodes correspond
to the real roots of  the polynomial (\ref{eq:nodal}). 
At Subsection D we present numerical results of the position of the nodes.

\begin{table}[]
\centering
\caption{First ten energy corrections $E^{(k)}_{n,l}$ for the Yukawa potential
in PT in powers of $\alpha$ for some selected states $(n,l)$.}
\label{Pcoeff}
\resizebox{\textwidth}{!}{
\begin{tabular}{l|ccccc}
\hline\hline
       $E^{(k)}$&$(1,0)$&$(2,0)$  &$(2,1)$  &$(3,0)$  &$(3,1)$  \\
\hline
$E^{(0)}$  &-1/2     &-1/8     &-1/8                &-1/18               &-1/18                             \\
$E^{(1)}$  &1     & 1    & 1                &  1             &  1                           \\
$E^{(2)}$  & -3/4     & -3   & -5/2               &  -27/4        &  -25/4                                   \\
$E^{(3)}$   & 1/2  & 7                & 5               &  69/2             &  30                                      \\
$E^{(4)}$ &-11/16 & -121/4            & -95/4            &  -5049/16      &  -2295/8                                 \\
$E^{(5)}$ &21/16 & 186               & 144              &  65043/16      &  29403/8                                 \\
$E^{(6)}$ &-145/48 & -8239/6           & -6431/6          &  -994437/16   &  -449307/8                               \\
$E^{(7)}$ &757/96 & 34414/3           & 26570/3          &  34182081/32   &  7672725/8                               \\
$E^{(8)}$ &-69433/3072 & -1256135/12        & -959575/12        &  -20438702541/1024 & -9115776855/512                       \\
$E^{(9)}$ &321499/4608 & 9197837/9        & 6926485/9       &  203591436363/512   & 45060827715/128                        \\
$E^{(10)}$&\quad-2343967/10240\quad &\quad-157991444/15\quad    &\quad -117213974/15\quad    & \quad-85364162187201/10240\quad     & \quad-37495774897443/5120\quad                     \\
\hline\hline
\end{tabular}}

\end{table}

\subsection{Critical Screening Parameter $\alpha_{c}^{(n,l)}$}

We carried out calculations of the energy of the low-lying states of the Yukawa potential 
using PT (+ Pad\'e approximants) and
Lag-Mesh. For the Lag-Mesh calculations, we wrote
a computational code in Fortran 90. The roots $r_{i}$, $i=1,2,..,N$, that define the mesh were calculated with \textit{Mathematica}. We diagonalized the matrix representation of  the secular equation (\ref{secular})   using the  DSYEV routine of LAPACK \cite{Diagonal}.

Far enough from $\alpha_c^{(n,l)}$, the Pad\'e approximants $P^{25}_{25}(\alpha)$ provide high
accuracy estimates of the energy $E_{n,l}$. As regards to the Lag-Mesh, 
a dimension $N\sim30$  generates accurate results that agree with those of  PT.

As the parameter $\alpha$ approximates to $\alpha_c^{(n,l)}\,$ 
a remarkable difference between  the energy $E_{n,l}$ obtained by the two methods appears.
Near  $\alpha_c^{(n,l)}$   the state $(n,l)$
becomes weakly bound, the corresponding wave function is very flat  and extended: an extension of the 
configuration domain of the variable $r$ is required. In PT 
an extension of the domain is achieved by calculating higher order terms in the expansion of the energy (\ref{eq:E}),  while in the Lag-Mesh we need
a larger dimension of the  mesh, i.e. a larger value of $N$. 
It is found that   the first 100 coefficients in the expansion (\ref{eq:E}) provide, via Pad\'e approximants, high accuracy for states with $l=0$ near $\alpha_{c}^{(n,0)}$. Otherwise, when $l\neq0$, even   $M=400$ in (\ref{eq:pades}) is not large enough to reach the accuracy of the Lag-Mesh. In any case, we only present results  with  diagonal Pad\'e approximants of the form $P^{200}_{200}(\alpha)$. 

For the results of Lag-Mesh we present only stable digits with respect to  variations of the dimension $N$. The largest value of $N$
considered was $N=2000$. In the case of $l=0$ we scale  $r$ with a positive parameter $g$ in the form   $r\rightarrow gr$. The parameter $g$
is such that the first positive energy is minimal.

The calculations of the energy are presented in Tables \ref{energy1s} - \ref{moreenergies}. In Table \ref{energy1s} 
we compare the energy of the ground state obtained by the two methods. One can  see that the difference appears in 
the 15th decimal digit. The same difference in digits appears in the  energy of  the excited states. In Table \ref{moreenergies} we  only present  the digits that are in agreement using both methods.  
In general, for $\alpha$ sufficiently far from the critical value $\alpha_c^{(n,l)}$, both methods  provides at least 14 decimal digits.
	\begin{table}[H]
		\centering
		\caption{Energy of the ground state ($1,0$) of the Yukawa potential as a function of $\alpha$. We present the results that come from PT and those from Lag-Mesh.}
		\label{energy1s}
	\begin{tabular}{lcclcc}
	\hline\hline
	\multicolumn{6}{c}{$-E_{1,0}$} \\
	\hline
	  $\alpha$&PT&Lag-Mesh.&$\alpha$&PT&Lag-Mesh\\
	  \hline
	  0.1&$\ 0.4070580306134030\ $   &$\ 0.4070580306134029\ $   &0.9   &$\ 0.0243141938275020\ $   &$\ 0.0243141938274960\ $  \\
	  0.2&0.3268085113691935   &0.3268085113691779   &1.0   &0.0102857899900177   &0.0102857899900174  \\
	  0.3&0.2576385863030541   &0.2576385863030501   &1.12   &0.0013846277112477   &0.0013846277112464  \\
	  0.4&0.1983760833618501   &0.1983760833618504   &1.14   &0.0007091358638104   &0.0007091358638094  \\
	  0.5&0.1481170218899326   &0.1983760833618504   &1.16   &0.0002586220063766   &0.0002586220063757  \\
	  0.6&0.1061359075058142   &0.1061359075058019   &1.18   &0.0000309859108740   &0.0000309859108732  \\
	  0.7&0.0718335559045121   &0.0718335559045112   &1.19   &0.0000001030319614   &0.0000000776158087  \\
	  0.8&0.0447043044973596   &0.0447043044973624   &   &   &  \\
	  	\hline\hline
	\end{tabular}

	\end{table}

 \begin{turnpage}
 \small
 \begin{table}[H]
 \centering
 \caption{Energy $E_{n,l}$ of some low-lying states as a function of $\alpha$. We present states with quantum numbers $n=2,3\,$ and $l=0,1,2\,$.}
 \label{moreenergies}
 \begin{tabular}{lllllllllllllllllllllll}
 \hline
 \hline
  $\quad\alpha$&&$\quad\quad\quad-E_{2,0}$& & &\quad$\alpha$ & &$\quad\quad\quad-E_{2,1}$& & &\quad$\alpha$ & &$\quad\quad\quad-E_{3,0}$& & &\quad $\alpha$  &&\quad\quad\quad$-E_{3,1}$& & &\quad $\alpha$ & & \quad\quad\quad$-E_{3,2}$  \\
 \hline
 0.001 & & 0.12400299303006 & &  & 0.001 & & 0.12400249502360 & & &  0.001 & & 0.05456227136711 & & &  0.001 & & 0.05456177583881 & & & 0.001 & &  0.05456078476559  \\  
 0.005 & & 0.12007414334559 & &  & 0.005 & & 0.12006188940983 & & &  0.005 & & 0.05072017847317 & & &  0.005 & & 0.05070822417583 & & & 0.005 & &  0.05068430583285  \\   
 0.01  & & 0.11529328516799 & &  & 0.01  & & 0.11524522409056 & & &  0.01  & & 0.04619885779903 & & &  0.01  & & 0.04615310482916 & & & 0.01  & &  0.04606145416065  \\   
 0.02  & & 0.10614832024469 & &  & 0.05  & & 0.08074038703778 & & &  0.02  & & 0.03802001439301 & & &  0.02  & & 0.03785238920022 & & & 0.02  & &  0.03751512770068  \\   
 0.04  & & 0.08941463418515 & &  & 0.06  & & 0.07314961938586 & & &  0.03  & & 0.03088608377997 & & &  0.03  & & 0.03054096758451 & & & 0.03  & &  0.02984182966659  \\   
 0.06  & & 0.07457853441270 & &  & 0.07  & & 0.06594417699615 & & &  0.04  & & 0.02469226725768 & & &  0.04  & & 0.02413235361039 & & & 0.04  & &  0.02298785675988  \\   
 0.08  & & 0.06146465621230 & &  & 0.08  & & 0.05911280478703 & & &  0.05  & & 0.01935255481475 & & &  0.05  & & 0.01855775188340 & & & 0.05  & &  0.01691557056981  \\   
 0.10  & & 0.04992827133191 & &  & 0.09  & & 0.05264570133158 & & &  0.06  & & 0.01479415729517 & & &  0.06  & & 0.01376134530350 & & & 0.06  & &  0.01160182947416  \\   
 0.12  & & 0.03984659244361 & &  & 0.10  & & 0.04653439048672 & & &  0.07  & & 0.01095392247489 & & &  0.07  & & 0.00969759375197 & & & 0.07  & &  0.00703987880545  \\   
 0.14  & & 0.03111313315223 & &  & 0.12  & & 0.03535143759661 & & &  0.08  & & 0.00777587703895 & & &  0.08  & & 0.00632999543926 & & & 0.08  & &  0.00324836046238  \\   
 0.15  & & 0.02722219072568 & &  & 0.14  & & 0.02552081310910 & & &  0.09  & & 0.00520944042038 & & &  0.09  & & 0.00363154381363 & & & 0.09  & &  0.00033545763623  \\   
 0.20  & & 0.01210786519544 & &  & 0.16  & & 0.01702093755237 & & &  0.10  & & 0.00320804674469 & & &  0.10  & & 0.00158900152586 & & & 0.091 & &  0.00001  \\   
 0.25  & & 0.00339590628323 & &  & 0.18  & & 0.00985980857128 & & &  0.12  & & 0.00072747319105 & & &  0.11  & & 0.000226339&         &                        \\ 
 0.30  & & 0.00009160244389 & &  & 0.20  & & 0.00410164653054 & & &  0.13  & & 0.00016543177793 & & &        & &&                      & \\ 
 0.31  & & 0.0000000379     & &  & 0.22  & & 0.00002          & & &  0.139 & & 0.0000004                &   & &                                                        \\ 
 \hline
 \hline
 \end{tabular}
 \end{table}
 \end{turnpage}

The critical values of  $\alpha_c^{(n,l)}$  are presented in Table \ref{criticos}. 
The results are compared with those of \cite{matrix}.
For states with angular momentum $l=0$, we find that Pad\'e approximants give more accurate
estimates of $\alpha_c^{(n,0)}$ in comparison with those of
Lag-Mesh (see results of  \cite{matrix}).
 Otherwise, when $l\neq0$, the 
Lag-Mesh provides more  accurate results in comparison with those of PT. 
As one can see in Table \ref{criticos}, the critical parameters obtained with Lag-Mesh exhibit more precision
as the angular momentum $l$ increases.

\begin{table}[H]
	\caption{\label{criticos} Critical parameter $\alpha_c^{(n,l)}$ of some low-lying states with 
		principal quantum number $n=1,2,3,4,5$ and  angular momentum $l=0,1,2$. 
		The results marked with $*$ in the first row were obtained with Lag-Mesh, the results
		marked with $**$ are the results of PT (+ Pad\'e approximants) and the results marked with $\dagger$ were obtained in
		\cite{matrix}.
	}
\resizebox{\textwidth}{!}{\begin{tabular}{|c|l|l|l|l|l|}

\hline
\hline
\diagbox{$n$}{$l$} & \quad\quad\quad\quad\quad0 & \quad\quad\quad\ 1 & \quad\quad\quad\quad2 &\quad\quad\quad\quad 3 &\quad\quad\quad\quad \,\,4\\
\hline
   & $1.190\,612^*$ & & & &\\
  1&$1.190\,612\,421\,060\,617\,70^{**}$&& & &\\
   & $ 1.190\,612\,421\,060\,618^{\dagger}$ & & &  & \\
   \hline
   & $0.310\,209^* $ & $0.220\,216\,806\,605^*$ & & &\\
  2&$0.310\,209\,282\,713\,935\,93^{**}$&$0.220\,216\,275^{**}$& && \\
   & $ 0.310\,209\,282\,713\,937^{\dagger}$ &$0.220\,216\,806\,61^{\dagger}$ & & &\\
   \hline
   &$ 0.139\,450^*$ & $0.112\,710\,498\,359^*$ & $0.091\,345\,120\,771\,732^*$ & & \\
  3&$0.139\,450\,294\,064\,178\,01^{**}$&$0.112\,708\,652\,7^{**}$&$0.091\,345\,116$ & &\\
   & $0.139\,450\,294\,064\,18^{\dagger}$ &$0.112\,710\,498\,36^{\dagger}$ &$0.091\,345\,120\,771\,732^{\dagger}$ & & \\
   \hline
   & $0.078\,428^*$ &$0.067\,885\,376\,100^*$ & $0.058\,105\,052\,754\,469^*$ & $0.048\,731\,132\,318\,646^{*}$ & \\
  4&$0.078\,828\,110\,273\,170\,6^{**}$&\quad\quad\quad\ \ -&\quad\quad\quad\quad\ - &\quad\quad\quad\quad\ -& \\
   & $ 0.078\,828\,110\,273\,172^{\dagger}$ &$0.067\,885\,376\,10^{\dagger}$ &$0.058\,105\,052\,754\,469^{\dagger}$
   & $0.049\,831\,132\,318\,646^{\dagger} $ & \\
   \hline
   & $0.050\,583^*$ &$0.045\,186\,248\,071^*$ & $0.040\,024\,353\,938\,324^*$ & $0.035\,389\,389\,799\,949^{*}$ &
   $0.031\,343\,552\,436\,537^{*}$\\
  5&$0.050\,583\,170\,374\,5^{**}$&\quad\quad\quad\ \ -&\quad\quad\quad\quad\ -&\quad\quad\quad\quad\ -&\quad\quad\quad\quad\ -\\
   & $ 0.050\,583\,170\,560^{\dagger}$ &$0.045\,186\,248^{\dagger}$ &$0.040\,024\,353\,938\,325^{\dagger}$
   & $0.035\,389\,389\, 799\,949$ & \quad\quad\quad\quad\ -\\
\hline \hline
\end{tabular}}
\end{table}

\subsection{Expansion of the Energy   at $\alpha_c^{(n,l)}$}

Near the critical parameter $\alpha_c^{(n,l)}$, the energy 
of the states $(n,0)$ can be represented as an expansion
(\ref{expansionE}), while the energy of the states ($n,l\neq0$)
is represented as an expansion of the type (\ref{expansionE2}).
Interestingly, both expansions (\ref{expansionE}) and  (\ref{expansionE2}) can be re-expanded at $\alpha=\alpha_c^{(n,l)}$ to  obtain a 
Taylor series and a Puiseux series in half-integer powers, respectively,
\begin{equation}
\label{expansionE*}
E_{n,0}(\alpha)\ =\ \sum_{k=2}^{\infty}\tilde{\beta}_k^{(n,0)}\left(\alpha_c^{(n,0)}-\alpha\right)^{k}\ 
,\quad\quad\quad E_{n,l\neq0}(\alpha)\ =\ \sum_{k=2}^{\infty}\tilde{\beta}_{k-1}^{(n,l)}\left(\alpha_c^{(n,l)}-\alpha\right)^{k/2}\ .
\end{equation} 
The coefficients $\beta_k^{(n,l)}$ of (\ref{expansionE}) and  (\ref{expansionE2}) 
are related with the coefficients $\tilde{\beta}_k^{(n,l)}$ of (\ref{expansionE*}) as follows
\begin{eqnarray}
 \label{betatildel0}
  l \neq  &0& \rightarrow \tilde{\beta}_2^{(n,l)}=\beta_2^{(n,l)}, 
  \quad \tilde{\beta}_3^{(n,l)}=\frac{\beta_3^{(n,l)}}{ \alpha_c^{(n,l)}}, \nonumber \\ 
 l \neq &0& \rightarrow \tilde{\beta}_1^{(n,l)}=\beta_1^{(n,l)}\alpha_c^{(n,l)},\quad
 \tilde{\beta}_2^{(n,l)}=\beta_2^{(n,l)}\sqrt{\alpha_c^{(n,l)}},\quad 
 \tilde{\beta}_3^{(n,l)}=\beta_3^{(n,l)}-\beta_1^{(n,l)}\,,
\end{eqnarray}
therefore we can calculate either $\beta_k^{(n,l)}$ or $\tilde{\beta}_k^{(n,l)}$.

We calculated the coefficients $\beta_k^{(n,l)}$ using the interpolation 
procedure mentioned in the Section I.
Alternatively the first coefficients $\tilde{\beta}_k^{(n,l)}$ can be 
found using the Hellman-Feynman, via expectation values  
 \begin{eqnarray}
 \label{beta1}
 \tilde{\beta}_2^{(n,0)} &=&\frac{1}{2}\,\partial_{\alpha}^2E_{n,0}\Bigr|_{\alpha=\alpha_c^{(n,0)}}=\frac{1}{2}\ \expval{\partial^2_\alpha\hat{h}}{\Psi_{n,0}}\ +\ \bra{\partial_\alpha\Psi_{n,0}}\partial_\alpha\hat{h}\ket{\Psi_{n,0}}\Bigr|_{\alpha=\alpha_c^{(n,0)}}\ ,  \\
 \tilde{\beta}_1^{(n,l)} &=&-\partial_{\alpha}E_{n,l}\Bigr|_{\alpha=\alpha_c^{(n,l)}}=\ -\expval{\partial_\alpha\hat{h}}{\Psi_{n,l}}\Bigr|_{\alpha=\alpha_c^{(n,l)}}\,,
   \label{beta2}
 \end{eqnarray} 
 where $\hat{h}$ represents the radial Hamiltonian  (\ref{eq:radialoperator}).
 However the Lag-Mesh wave function (\ref{eq:ulag}) does not contain explicit dependence 
  on $\alpha$ and therefore  it is only applied straightforwardly to obtain $\tilde{\beta}_{1}^{(n,l)}$, see (\ref{beta1}). Using the expectation value of a function within the Lag-Mesh via (\ref{matrixV})
  we estimated the coefficient $\tilde{\beta}_{1}^{(n,l)}$.

  Pad\'e approximants  turn out to be appropriate to calculate straightforwardly the coefficients  $\tilde{\beta}_{n,0}$ by means of its Taylor series at $\alpha=\alpha_c^{(n,l)}$. Consequently no interpolation for states ($n,0$) is needed to construct the corresponding expansion presented in (\ref{expansionE*}). However, for states with $l\neq0$
  we found that the estimation of the  coefficient $\tilde{\beta}_{1}^{(n,l\neq0)}$ by 
  means of the Pad\'e approximants  
 $P^{ \left \lceil{M/2}\right \rceil }_{\left \lfloor{M/2}\right \rfloor}(\alpha)$ converges too slowly as $M$ increases (even $M=400$ is not large enough).
 
 The first three dominant coefficients are presented in Table \ref{betas}.
 It must be remarked that the coefficients $\tilde{\beta}_{i=1,2,3}^{(n,l)}$ are always negative and 
  decreasing as a function of $n$  (for a fixed angular momentum $l$).
 Since  the critical parameter $\alpha_c^{(n,0)}$ provided by PT is more accurate than that of Lag-Mesh, we expect that PT provides more  accurate coefficients $\tilde{\beta}_{i=2,3,...}^{(n,0)}$. 
 Otherwise (if $l\neq 0$) the coefficient $\tilde{\beta}_{1}^{(n,l)}$ provided by Lag-Mesh is more accurate 
 than that of PT.
 
 \begin{table}[H]
 	\caption{\label{betas}  First coefficients $\tilde{\beta}_k$ of the expansion of the energy in the neighborhood of the critical 
	parameter,  see (\ref{expansionE*}). We consider the low-lying states with 
	principal quantum number $n=1,2,3,4,5$ and  angular momentum $l=0,1,2$. 
	The results of PT are marked with the super index $*$, the results of Lag-Mesh with $**$ and 
	the results of the interpolating function with $\dagger$.
		} 
  \begin{center}
		\begin{tabular}{|l|lll|l|lll|} 
		
		\hline
			$(n,l)$ & \quad$\tilde{\beta}_1$ & \quad\quad$\tilde{\beta}_2$ &\quad\quad $\tilde{\beta}_3$ & $(n,l)$ & 
			\quad\ $\tilde{\beta}_1$ & \quad\ $\tilde{\beta}_2$ &\quad\ \  $\tilde{\beta}_3$ \\ \hline
			$(1,0)$& \quad\ - & -$0.274683^{*}$   & -$0.041881^{*}$ & $(3,2)$& -$0.221946^{*}$ & & \\ 
			$(2,0)$ & \quad\ - & -$0.867186^{*}$  & -$1.141192^{*}$ & & -$0.221917^{**}$ & & \\ 
			 $(2,1)$& -$0.1292^{*}$ &  & & & -$0.221916^{\dagger}$ & -$0.000426^{\dagger}$ & -$10.844400^{\dagger}$ \\
			  & -$0.1264^{**}$  & & & $(4,0)$&\quad\quad - & -$3.019744^{*}$  & $-27.462522^{*}$\\
			   & -$0.1260^{\dagger}$  & -$0.4182^{\dagger}$ & -$0.8856^{\dagger}$ & $(4,1)$& -$0.0369^{**}$ & &  \\ 
			 $(3,0)$& \quad\ - &-$1.783905^{*}$  & -$7.365419^{*}$ &  & -$0.0363^{\dagger}$ & -$0.3685^{\dagger}$ & -$2.3894^{\dagger}$ \\ 
			 $(3,1)$& -$0.0624^{*}$ & & & $(4,2)$& -$0.129012^{**}$ & &\\ 
			  & -$0.0597^{**}$ & & & & -$0.129011^{\dagger}$ & -$0.000597^{\dagger}$ & -$16.126500^{\dagger}$\\
			  & -$0.0592^{\dagger}$ & -$0.3799^{\dagger}$ & -$1.5677^{\dagger}$ &&&&\\ 
			\hline
		\end{tabular}
	\end{center}
\end{table}

\subsection{Nodes}

Using Pad\'e approximants of series (\ref{eq:coeff}) and the Lag-Mesh we  computed the real radial nodes of the states (2,0), (3,0) and (3,1). 
Results are presented in the Table \ref{nodals}. Just as for the energy estimates, the agreement between both approaches is 13 decimal digits if  $\alpha$ 
is sufficiently far from $\alpha_c^{(n,l)}$.  The results indicate that Pad\'e approximants of series  (\ref{eq:coeff}) converge to the exact result.  It is worth mentioning that  the radial nodes of any state do not exhibit any critical behavior near $\alpha_c^{(n,l)}$. They turn out to be continuous and smooth functions of $\alpha$, this fact is also reflected in coefficients $A_{n,l}[k]$, see (\ref{eq:coeff}). In Figures \ref{fig:2,0} - \ref{fig:3,1} we  plot of the coefficients $A_{n,l}[k]$ with $k=0,1,...,n-l-1$ as  functions of $\alpha$.
Interestingly, the coefficient $A_{n,l}[k]$  is a monotonic  function with a well-defined sign.
\begin{table}[H]
\centering
\caption{Radial nodes of some selected states $(n,l)$ as functions of $\alpha$ . We present only the digits which coincide in both methods 
PT (+ Pad\'e approximants) and Lag-Mesh. The state $(3,1)$ has a  node at $r=0$ for any value of $\alpha$. The numbers presented in the second row corresponds to the second finite node of the wave function of the state (3,0).
}
\label{nodals}
\begin{tabular}{llllllllllllllll}
\hline
\hline
     $\quad \alpha$&\quad\quad \quad\quad         $(2,0)$      & &  &  
     &\quad $\alpha$&\quad\quad \quad       $(3,0)$        & &  & &\quad $\alpha$ &\quad\quad \quad $(3,1)$  \\
\hline 
       0.001 &   2.00000398939430& &  &  &  0.001  &  1.9019318839321& & &  & 0.001 &  6.0001073618376  \\ 
             &                    & &  &  &        &  7.0981888374472& & &  &       &                    \\ 
       0.01  &   2.00038991232397& &  &  &  0.01   &  1.9026977265017& & &  & 0.01  &  6.0102406162908  \\ 
             &                    & &  &  &        &  7.1087623332409& & &  &       &                    \\ 
       0.12  &   2.04685329041174& &  &  &  0.10   &  1.9612099323332& & &  & 0.05  &  6.2289083505411  \\                	                                                      
             &                    & &  &  &        &  8.0003411447325& & &  &       &                    \\ 
       0.14  &   2.06255289338718& &  &  &  0.11   &  1.9723796976230& & &  & 0.06  &  6.3269229450805  \\  
             &                    & &  &  &        &  8.2019699771969& & &  &       &                    \\ 
       0.15  &   2.07120228041780& &  &  &  0.12   &  1.9843916443692& & &  & 0.07  &  6.4444890892019  \\  
             &                    & &  &  &        &  8.4345171899446& & &  &       &                    \\ 
       0.20  &   2.12263667670952& &  &  &  0.13   &  1.9972337679448& & &  & 0.08  &  6.5841588244220  \\ 
             &                    & &  &  &        &  8.7042030005712& & &  &       &                     \\      
       0.25  &   2.18866874660131& &  &  &  0.139  &  2.0094908935673& & &  & 0.09  &  6.7497210992869   \\ 
             &                    & &  &  &        &  8.9856673091028& & &  &       &                     \\ 
       0.30  &   2.27150103691977& &  &  &         &                  & & &  & 0.10 &  6.9469183539167   \\ 
             &                    & &  &  &        &                  & & &  &      &                     \\ 
       0.31  &   2.29036843836199& &  &  &         &                  & & &  & 0.112&  7.2400431333415   \\                                                                                               
\hline                                                                  \hline                   
\end{tabular}                                                                               
\end{table}

\begin{figure}[H]
\centering
\includegraphics[width=0.4\textwidth]{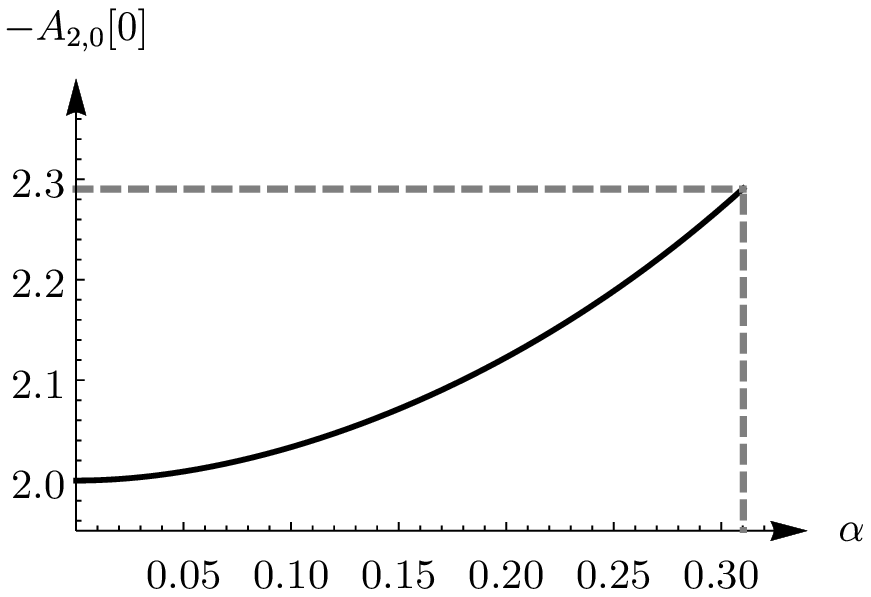}
        \caption{Absolute value of  $A_{2,0}[0]$ (calculated by means of Pad\'e approximants) as a function of $\alpha$. The dashed lines indicates the position of $\alpha_c^{(2,0)}$ and the corresponding value of the coefficient $A_{2,0}[0]$.}
        \label{fig:2,0}
\end{figure}
\begin{figure}[H]
    \centering
    \begin{minipage}{0.5\textwidth}
        \centering
        \includegraphics[width=0.85\textwidth]{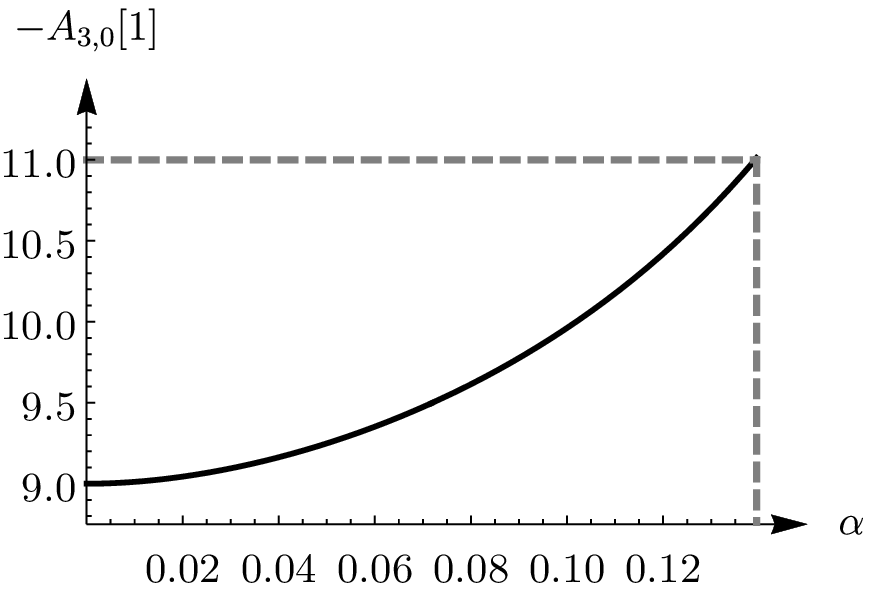} 
    \end{minipage}\hfill
    \begin{minipage}{0.5\textwidth}
        \centering
        \includegraphics[width=0.85\textwidth]{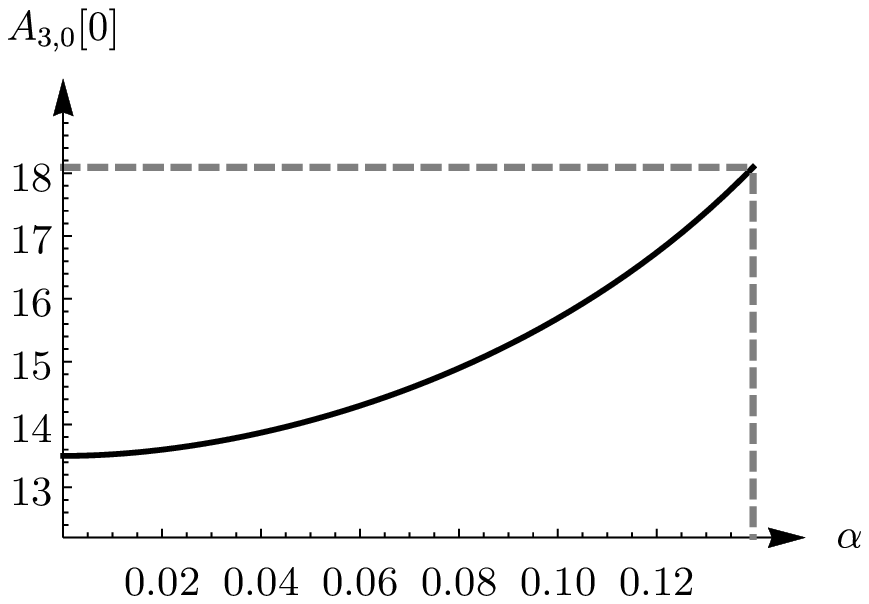} 
    \end{minipage}
\caption{Absolute value of  $A_{3,0}[1]$  and  $A_{3,0}[0]$ (both calculated by means of Pad\'e approximants) as a function of $\alpha$. The dashed lines indicates the position of $\alpha_c^{(3,0)}$ and the corresponding value of the coefficient $-A_{3,0}[1]$ and $A_{3,0}[0]$.}
\end{figure}
\begin{figure}[H]
\centering
\includegraphics[width=0.4\textwidth]{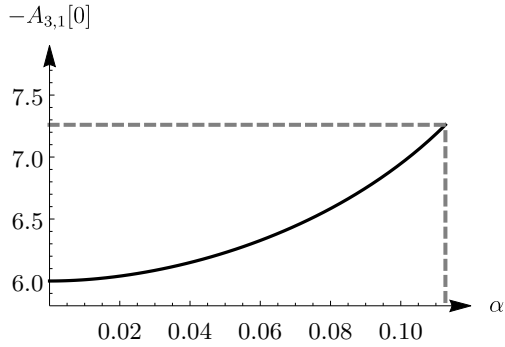}
        \caption{Absolute value of  $A_{3,1}[0]$ (calculated by means of Pad\'e approximants) as a function of $\alpha$. The dashed lines indicates the position of $\alpha_c^{(3,1)}$ and the corresponding value of the coefficient $A_{2,0}[0]$. }
        \label{fig:3,1}
\end{figure}

\subsection{Variational Calculation and its Accuracy}
We carried out variational calculations for  the  states with  quantum numbers 
$(1,0)$, $(2,0)$ and $(2,1)$. A computational code was written in FORTRAN 90 in other to perform variational calculations. The integration routine was designed using the algorithm  described in \cite{Integracion}. The optimization of the variational parameters was performed by means of the MINUIT routine \cite{Minuit} of CERN-LIB.

In Table \ref{table:var} we present variational energies of the states (1,0) and  (2,1)
for different values of $\alpha$. We also present their corresponding corrections and approximations (\ref{notation}) up 
to third order in PT. In Table \ref{table:varbis} we present the variational energy of the state (2,0).
For this state we include up to the second order corrections and the corresponding approximations of the energy.
Additionally, Table \ref{table:varbis} contains the zeroth and first order approximation of the position of the radial node.

We must emphasize that the nodal surface of the exact  wave functions of 
the states  $(1,0)$ and ($2,1$) do not depend on $\alpha$. 
Of course this information is codified in their corresponding trial 
functions $\Psi_{1,0}^{(t)}$ and $\Psi_{2,0}^{(t)}$, see (\ref{tfes}). Therefore, 
no correction to the nodal surface has to be calculated. In contrast, the state (2,1) 
develops a node at finite $r\neq0$ which can be only approximated by numerical means. For this state, following (\ref{tfes}), 
the trial function reads 
\begin{equation}
\Psi_{2,0}^{(t)}(r)\ =\ \left(r-r_0^{(0)}\right)\,e^{-\Phi_{2,0}^{(t)}}\,,
\end{equation}
 where $r_0^{(0)}$ is the zeroth order approximation of the position of the node. 
 In fact $r_0^{(0)}$ is determined from the orthogonality constraint that we impose between 
 the trial functions of the states (1,0) and (2,0), 
 \begin{equation}
 r_0^{(0)}\ =\ \frac{\int_{0}^{\infty}e^{\Phi_{1,0}^{(t)}}\,e^{-\Phi_{2,0}^{(t)}}\,r^3\,dr}{\int_0^{\infty}e^{\Phi_{1,0}^{(t)}}\,e^{-\Phi_{2,0}^{(t)}}\,r^2\,dr}\ .
 \end{equation} 
The first order  correction of $r_0^{(0)}$  must be a constant and, 
from equation (\ref{eq:kcorrection}) together with boundary conditions (\ref{boundary}), one can show that it is given by
\begin{equation}
r_0^{(1)}\ =\ \frac{2\int_{0}^{r_0^{(0)}} (E_{2,0;1}-V^{(1)})\,\left(\Psi_{2,0}^{(t)}\right)^{2}\,r^2\,dr}{\left(r_0^{(0)}\right)^2\,e^{-2\Phi_{2,0}^{(t)}}}\ .
\end{equation}
Higher order corrections $r_0^{(k)}$ with  $k>1$ can be calculated, however, the higher order correction, the more cumbersome calculations have to be performed.
As one can expect, the calculation of the corrections  in PT is simpler if the exact function $f_{n,l}(r)$ is known.

 \begin{table}[H]
 \centering
 \caption{Variational energy of the state $(1,0)$ and $(2,1)$ for some selected values of $\alpha$. We present  up to the third order correction in PT. For corrections and approximations of different order we take the notation 
 used in (\ref{notation}). The digits presented for $E_{1,0;3}$ correspond to digits verified with Pad\'e approximants and the Lag-Mesh. }
   
 \resizebox{\textwidth}{!}{\begin{tabular}{clllll}
 \hline
 \hline
  $\alpha$ & \quad\quad\quad-$E_{1,0;1}$  &\quad\quad$E_{1,0}^{(2)}$  &\quad\quad\quad-$E_{1,0;2}$  &$\quad\quad E_{1,0}^{(3)}$  &\quad\quad\quad-$E_{1,0;3}$  \\
  \hline
  0.01&$0.4900745066986851$ &-$4.80011\times10^{-11}$&$0.4900745067466863$&-$7.97\times10^{-15}$&$0.4900745067466942$ \\

     0.10&$0.40705803060984105$&-$3.56285\times 10^{-12}$&$0.40705803061340390$&$7.58\times10^{-16}$&0.40705803061340315\\
     0.50&$0.148117017843659$&-$4.046202\times10^{-9}$&$0.148117021889861$&-$7.13\times 10^{-14}$&$0.148117021889933$\\
     1.00&$0.01028575123218$&-$3.874918\times10^{-8}$&$0.01028578998128$&-$8.7607\times10^{-12}$&$0.01028578999004$\\
      \hline
     $\alpha$ & \quad\quad\quad-$E_{2,1;1}$  &\quad\quad$E_{2,1}^{(2)}$  &\quad\quad\quad-$E_{2,1;2}$  &\quad\quad$E_{2,1}^{(3)}$  &\quad\quad\quad-$E_{2,1;3}$  \\
     \hline
     0.01&0.115245224090557422&-$6.762\times10^{-15}$&0.115245224090564185&-$3.7\times10^{-20}$&0.115245224090564185\\
     0.10&0.046534388129&-$2.341\times10^{-9}$   &0.046534390471&-$1.52\times10^{-11}$&0.046534390486\\
     0.15&0.021104690&-$1.924\times10^{-7}$&0.021104882&-$5.80\times10^{-9}$&0.021104888\\
     0.20&0.004093&-$6.84\times10^{-6}$&0.004100&-$7.16\times10^{-7}$&0.004101\\
     \hline
     \hline
 
 \end{tabular}}
 \label{table:var}
 \end{table}

 \begin{table}[H]
	\centering
	\caption{Variational energy of the first state (2,0) and its corrections in PT for some selected values of $\alpha$. For corrections and approximations of different order we take the notation 
	 used in (\ref{notation}). The digits presented for $E_{1,0;2}$ and $r_0^{(0)}+r_0^{(1)}$ correspond to digits verified with Pad\'e approximants and the Lag-Mesh.}
	
	\resizebox{\textwidth}{!}{\begin{tabular}{lllllll}
		\hline
		\hline
		\quad$\alpha$ & \quad\quad-$E_{2,0;1}$  &$\quad\quad E_{2,0}^{(2)}$  &\quad\quad-$E_{2,0;2}$&$\quad\quad r_0^{(0)}$&$\quad\quad r_{0}^{(1)}$&\quad$r_0^{(0)}+r_0^{(1)}$   \\
		\hline
		0.01\quad\quad&0.11529328517404\quad\quad&$6.05\times10^{-12}\quad$&0.11529328516799\quad\quad&2.000381392\quad\quad&-$8.519\times10^{-6}\quad$&2.000389912\\
		0.10&0.0499282672&-$4.05\times10^{-9}$&0.0499282713&2.03333727&$5.2755\times10^{-5}$&2.03328451\\
		0.2&0.012104&-$3.27\times10^{-6}$&0.012107&2.12385&0.00120&2.12266\\
		0.25&0.00336&-$2.48\times10^{-5}$&0.00339&2.189 &0.0008&2.188\\
		\hline
		\hline
 	\end{tabular}}
	 \label{table:varbis}
\end{table}
Results indicates that  PT developed for the variational  energy  leads to a  fast convergent series. Therefore, equation (\ref{eq:limit}) should be fulfilled. In the case of the ground state (1,0) we are able to provide 14 - 17 exact digits in the range $0\leq\alpha\leq 1$.
For the three states studied,  the   variational calculation of energy becomes less accuracy as $\alpha\rar\alpha_c$. In general, the trail function with optimized parameters becomes very flat. In this regime, by calculating corrections of the energy we obtain a slow convergent series. 
For the state (2,0) we obtain also a high accurate value for the position of the node which agree with previous results presented in Table \ref{nodals}. 

The Non-Linearization Procedure allows us to estimate the absolute deviation between  the exact wave function $\Psi_{n,l}$ and the trial function $\Psi_{n,l}^{(t)}$. For example,  for the ground state  we obtain a very small deviation 
\begin{equation}
 |\Psi_{1,0}(r)-\Psi_{1,0}^{(t)}(r)|<1\times10^{-4}\ ,
\end{equation}
in the  domain $0\leq r<\infty$ with $0\leq\alpha<1$. In this sense, we conclude that the trial wave function $\Psi_{1,0}^{(t)}$ that we designed is locally accurate. For the states (2,0) and (2,1) the same situation  occurs.

\subsection{Analytic Expression for $E_{n,l}$}

Following the prescription described in \cite{interpolaciones} we construct analytical expressions of the energy $E_{n,l}$.
We use the expansion of the energy in the regime of PT  (\ref{eq:E})
and the expansion in the neighborhood of the critical parameter 
(\ref{expansionE},\ref{expansionE2}) in order to  construct a compact interpolating function. 

The interpolating function is a rational function with free parameters to be adjusted by least squares.
The expansions of the interpolating functions at
$\alpha\rightarrow 0$ and $\alpha\rightarrow \alpha_c^{(n,l)}$ must reproduce functionally the expansions (\ref{eq:E}),
(\ref{expansionE},\ref{expansionE2}).
In order to do this, four restrictions are introduced such that only the dominant and subdominant 
terms  of each expansion are exactly reproduced. 

For the states with angular momentum $l=0$ we propose the function
\begin{equation}
 \label{PadeYukawal0}
 E_{n,0}(\alpha)=\frac{\sum_{i=2}^M c_i^{(n,0)}(\alpha_c^{(n,0)}-\alpha)^i}{\sum_{j=0}^N d_j^{(n,0)}(\alpha_c^{(n,0)}-\alpha)^j}\,,
\end{equation}
where $c_i^{(n,0)}$ and $d_j^{(n,0)}$ are parameters. The parameter $d_0^{(n,0)}$ is fixed to the unity as a normalization.
In this case the function contains $M+N-5$ free parameters.
For states with $l\neq 0$ we propose the function
\begin{equation}
 \label{PadeYukawal}
 E_{n,l}(\alpha)=\frac{\sum_{i=2}^M c_i^{(n,l)}(\alpha_c^{(n,l)}-\alpha)^{i/2}}{\sum_{j=0}^N d_j^{(n,l)}(\alpha_c^{(n,l)}-\alpha)^j}\,,
\end{equation}
where $c_i^{(n,l)}$ and $d_j^{(n,l)}$ are parameters and $d_0^{(n,l)}$ is fixed to the unity.
In this case the interpolating function also contains $M+N-5$ free parameters.

For all cases,  $M=N=4$  is the the minimal value which provides at least 
$6$ correct digits for any value of $\alpha$.  In Table \ref{table:varnodes} we present the fitted parameters of the states (1,0), (2,0) and (2,1).
\begin{table}[H]
	\centering
	\caption{Fitted parameters of the interpolating functions (\ref{PadeYukawal0}) and (\ref{PadeYukawal}) for some states $(n,l)$.}
	\begin{tabular}{l|lllllll}
		\hline
		\hline
		$(n,l)$&$c_2^{(n,l)}$  &$c_3^{(n,l)}$ &$c_4^{(n,l)}$& $d_1^{(n,l)}$&$d_2^{(n,l)}$ &$d_3^{(n,l)}$ &$d_4^{(n,l)}$  \\
		\hline
		(1,0)&-0.274683&0.273223  &-0.064690  &-1.147158  &0.369868  &-0.0228517  &-0.001716  \\
		(2,0)&-0.867186  &0.297361  &2.965554&-1.658874&-1.40866   &0.481608   &1.367597      \\
		(2,1)&-0.126016&-0.418180& -0.840617&-0.0330058&-3.268627  &9.004197  &-15.528311 \\
		\hline
		\hline
	\end{tabular}
	\label{table:varnodes}
\end{table}

 \section{Conclusions}
The Yukawa potential is studied using three different approaches: PT, Lag-Mesh and VM. Perturbation series 
in powers of $\alpha$ are calculated algebraically within the framework 
of the Non-Linerazation Procedure.
It is shown that the Pad\'e approximants related to the  PT series of
the energy provide  13 - 14 exact decimal digits
which are in agreement with the Lag-Mesh calculations. 
The methods of PT (+ Pad\'e approximants) and Lag-Mesh  describe 
correctly the behavior of a particle 
 in the Yukawa potential even near the critical parameters where the states become 
 weakly bound.
For all states considered, we reproduce or even exceed the precision of  the critical
screening parameters that are known in  the literature to the  authors at the moment.
The  nodes of the radial wave function are found for various states and 
also exhibit a good agreement between the three methods. 
The Pad\'e approximants allow us to calculate the nodes as the roots of some
polynomials.  
We design a locally accurate and compact trial function whose absolute deviation from 
the exact wave function is less than $10^{-4}$ in the range of $r$.
Together with perturbative 
corrections within
the Non-Linearization Procedure, the trial function leads to highly accurate estimates of the energy
comparable with  other numerical methods (in particular, Lag-Mesh). 
The knowledge of a highly accurate trial functions  will allow to calculate transition amplitudes
with high accuracy and hence to create a theory of radiative transitions for the Yukawa potential.
Finally, a remarkable analytical expression for the energy of several states 
is obtained: it  reproduces 6  significant digits correctly in the entire physical range of 
the screening parameter $\alpha$.

 \section*{Acknowledgments}

 The authors thank A. V. Turbiner for numerous discussions and 
 constructive suggestions,  H. Olivares-Pil\'on and J. C. L\'opez Vieyra,  both
 for the interest to the work and useful remarks. This work was supported by  CONACyT(Mexico) PhD. 
 scholarships.

\end{document}